\documentclass[12pt,a4paper]{article}
\usepackage{cite}
\usepackage{amsmath}
\usepackage{graphicx}

\begin{document}
\input feynman

\begin{center}
\LARGE{The Behavior of Form Factors of Nucleon Resonances and Quark-Hadron Duality}

\vspace{0.5cm}
\large{V.V. Davidovsky $^{*}$ \footnote[1]{e-mail: odavi@kinr.kiev.ua}, B.V. Struminsky $^{**}$}

\vspace{0.3cm}
\small{$^{*}$\it Institute for Nuclear Research, Kiev, Ukraine.}

\small{$^{**}$\it Bogolyubov Institute for Theoretical Physics, Kiev, Ukraine.}
\end{center}
\vspace{0.5cm}
\begin{abstract}
A behavior of the nucleon structure functions in resonance region
is investigated. Expressions for resonance production form
factors, dependent on photon virtuality $Q^2$, which have correct
threshold behavior and take into account the available data on
resonance decays, are obtained. Contributions of resonances to nucleon
structure functions are calculated. Obtained expressions are used to investigate
the quark-hadron duality in the electron-nucleon
scattering processes with the structure function $F_2$ as an example.
\end{abstract}

\section{Introduction}
Many years ago the duality of resonances at
low energies and Regge poles at high energies was found. The term
"duality" means here that an amplitude can be described either by
the resonances or by the reggeons.

Veneziano amplitude was the first example of successful
implementation of hadron-reggeon duality. Another example is a
Dual Amplitude with Mandelstam Analyticity (DAMA) \cite{Bugrij&2:NuclPhys:1971,
Jenkovszky&2:EPhJ:2001}, in which the complex non-linear Regge
trajectories with correct thresholds can be introduced.

A little bit later the quark-hadron duality was discovered by
Bloom and Gilman \cite{Bloom&1:PhysRev:1971}. They analyzed nucleon structure function $W_2$ and have
found that structure function in the nucleon resonance excitation
region, averaged over resonances, with good accuracy coincides with the structure function in the deep inelastic region.

The exact mathematical formulation of the duality (both resonance-reggeon and quark-hadron) was given on the
basis of the
finite-energy sum rules, formulated independently in the works
\cite{Igi&1:PhysRevLett:1967, Logunov&2:PhysLett:1967,Dolen&2:PhysRev:1968}.

The investigation of the quark-hadron duality can improve our understanding
of the structure and interaction of hadrons in terms of quark and gluon degrees of
freedom.

Quark and gluon degrees of freedom are convenient basis for  the
gauge invariant theory of strong interaction -- quantum
chromodynamics (QCD). Hadronic degrees of freedom form other basis.
Since all physical quantities must not depend on the choice of the
basis, the descriptions of processes in terms of quark-gluon and
hadronic degrees of freedom must be equivalent.

The choice of degrees of freedom for the description of hadron
processes depends on specific kinematic conditions.
For instance, in lepton-nucleon scattering the resonance production region is usually described
using hadronic degrees of freedom, whereas the deep inelastic region is naturally described
in terms of quark-gluon degrees of freedom.

Formally, the quark-hadron duality is exact, but practically
the necessity to cut the expansion of any Fock state leads to
different manifestations of the duality under various kinematic
conditions and in various reactions.

The duality offers additional possibilities for the investigation
of nucleon structure using the data on the properties of nucleon resonances.
For example, the corrections to scaling behavior of the structure
function $F_2$, calculated in QCD and measured during the last decade,
can be extracted from the resonance data \cite{Carlson&1:PhysRevLett:1995}.

For theoretical description of the duality it's necessary to construct the structure functions
in the resonance region (threshold energies and small photon virtualities).
For this purpose the dependence of resonance production form factors $\gamma^{*}N\to R$ on
photon virtuality $Q^2$ must be known. In this article we construct the form factors and study the quark-hadron
duality.

The experimental data, obtained in Jefferson LAB \cite{Niculescu&41:PhysRevLett:2000,Niculescu:PhDThesis:1999}, served as a good
incentive for this work.

\section{Form Factors}

Lepton-hadron scattering in the lowest order on electromagnetic
coupling constant is described as an exchange of virtual photon with virtuality
$Q^2=-q^2$ ($q$ is four-dimensional momentum of photon)
and energy (in nucleon rest frame) $\nu = (pq)/m$ ($p$ is four-dimensional momentum of nucleon,
$m$ is nucleon mass). In this case $q^2$ serves the role of squared momentum transferred from lepton to
nucleon.

The mechanism of virtual photon and nucleon interaction depends on
the quantity of transferred momentum $Q^2$ and photon energy.
At high $Q^2$ and $\nu$ noncoherent scattering of virtual photon on nucleon quarks takes place,
and huge amount of hadrons is produced.

At moderate $Q^2$ ($\sim 1$ GeV$^2$) and $\nu$
internal nucleon spin states are excited, which leads to the resonance production (fig. 1).

Below we introduce main notations and define necessary quantities.

Four-dimensional momentum of nucleon, photon and resonance are denoted as $p$,
$q$ and $P$ respectively. At the resonance rest frame components
of this vectors read:
\begin{gather}
p=(\frac{M^2+m^2+Q^2}{2 M}; 0, 0, -\frac{\sqrt{(M^2-m^2-Q^2)^2+4
M^2 Q^2}}{2 M}),\\ q=(\frac{M^2-m^2-Q^2}{2 M}; 0, 0,
\frac{\sqrt{(M^2-m^2-Q^2)^2+4 M^2 Q^2}}{2 M}),\\ P=(M; 0, 0, 0),
\end{gather}
where $p^2=m^2$, $P^{2}=M^{2}$ and $M$ is the mass of resonance.

The vertex of virtual photon absorption by nucleon $\gamma^{*} N\to
R$ is described by three independent form factors $G_{\pm,0}(Q^2)$
(or by two, when a spin of resonance equals to $1/2$, and
$G_{-}(Q^2)\equiv 0$), which are (in the resonance rest frame)helicity amplitudes of the transition
$\gamma^{*} N\to R$:
\begin{equation}\label{Gspiral}
G_{\lambda_\gamma}=\frac{1}{2m}<R,\lambda_R=\lambda_N-\lambda_\gamma|J(0)|N,\lambda_N>\,,
\end{equation}
where $\lambda_R$, $\lambda_N$ and $\lambda_\gamma$ are the helicity of
resonance, nucleon and photon respectively; $J(0)$ is current operator;
$\lambda_\gamma$ takes values of $-1$, $0$, $+1$.

Nucleon structure functions can be expressed in terms of form
factors (\ref{Gspiral}) as follows \cite{Carlson&1:PhysRev:1998}:
\begin{equation}\label{F1}
F_1(x,Q^2)=m^2\,\delta(W^2-M^2)\,\left[|G_{+}(Q^2)|^2+|G_{-}(Q^2)|^2\right]\,,
\end{equation}
\begin{equation}
\left(1+\frac{\nu^2}{Q^2}\right)\,F_2(x,Q^2)=m\,\nu\,\delta(W^2-M^2)\,\left[|G_{+}(Q^2)|^2+2|G_0(Q^2)|^2+|G_{-}(Q^2)|^2\right]\,,
\end{equation}
\begin{multline}
\left(1+\frac{Q^2}{\nu^2}\right)\,g_1(x,Q^2)=m^2\,\delta(W^2-M^2)\,\left[|G_{+}(Q^2)|^2-|G_{-}(Q^2)|^2+\right.
\\ \left. +
(-1)^{J-1/2}\,\eta\,\frac{Q\sqrt{2}}{\nu}\,G_0^{*}(Q^2)\,
G_{+}(Q^2)\right]\,,
\end{multline}
\begin{multline}\label{g2}
\left(1+\frac{Q^2}{\nu^2}\right)\,g_2(x,Q^2)=-m^2\,\delta(W^2-M^2)\,\left[|G_{+}(Q^2)|^2-|G_{-}(Q^2)|^2-\right.
\\ \left. -
(-1)^{J-1/2}\,\eta\,\frac{\nu\sqrt{2}}{Q}\,G_0^{*}(Q^2)\,
G_{+}(Q^2)\right]\,,
\end{multline}
where $J$, $\eta$ are the spin and parity of resonance respectively;
$W^2=(p+q)^2$ is the square of total energy of photon and nucleon
in the c.m. frame; $x$ is Bjorken variable. Just to remind, the
structure functions $F_{1}$ and $F_{2}$ are related to $W_{1}$ and $W_{2}$
by the relations $F_{1}(x,Q^2)= m W_{1}(\nu,Q^2)$ and $F_{2}(x,Q^2)=
\nu W_{2}(\nu,Q^2)$.

Now, it's easy to derive the relation between structure functions
$F_1$ and $F_2$:
\begin{equation}
(1+\frac{4m^2x^2}{Q^2})\,F_2=2xF_1\,(1+R(Q^2))\,.
\end{equation}
This relation in the scaling limit $Q^2\to \infty$ under
assumption $R(Q^2\to \infty)\to 0$ transforms to well known
Callan-Gross relation valid in parton model for quarks with spin
1/2.

Formulae (\ref{F1})--(\ref{g2}) determine a contribution of one infinitely narrow resonance
into nucleon structure functions. For a resonance with width $\Gamma$
in the expressions (\ref{F1})--(\ref{g2}) we change delta-function $\delta(W^2-M^2)$
to
\begin{equation}\label{resonance}
\frac{1}{\pi}\,\frac{M \Gamma}{(W^2-M^2)^2+M^2 \Gamma^2}\,.
\end{equation}
In principle, this expression is not unique approximation
of resonance shape. But now the particular choice of resonance contribution
is not very significant. It's worth mentioning here that the expression (\ref{resonance})
originates from the propagator of a resonance.

The basic idea of this paper is to take contributions of all resonances,
which data is published in literature \cite{PDG:PhysLett:1988}, into
account.
If we let $F^{R}_{1,2}$ and $g^{R}_{1,2}$ to denote the contribution of resonance $R$ into spin-independent and
spin-dependent structure function respectively, then the contribution of only resonances to the structure functions
will be written in the form of the sum:
\begin{equation}\label{StrFun_Sum}
F_{1,2}=\sum_{R}F^{R}_{1,2};\quad g_{1,2}=\sum_{R}g^{R}_{1,2}.
\end{equation}

To calculate resonance contribution to the structure function one must
construct the form factors of resonance production as functions of
photon virtuality $Q^2$. Note that experimentally the dependence of form factors of known resonances
\cite{PDG:PhysLett:1988} on $Q^2$ practically is not studied. In
tables \cite{PDG:PhysLett:1988} only their values at $Q^2=0$ are listed.

The transitions $\gamma^{*}N\to R$ by the parity may be of two types:
normal, i.e.
\begin{equation}\label{N}
1/2^{+}\to 3/2^{-}, 5/2^{+}, 7/2^{-}, ...
\end{equation}
and abnormal:
\begin{equation}\label{A}
1/2^{+}\to 1/2^{-}, 3/2^{+}, 5/2^{-}, ...
\end{equation}

About corresponding form factors $G_{\pm,0}(Q^2)$ it's known: a) form factor threshold behavior
at $|\vec{q}|\to 0$ \cite{Bjorken&1:AnnPhys:1966}, b) form factor asymptotic behavior
at high $Q^2$, c) form factor value at $Q^2=0$ \cite{PDG:PhysLett:1988}.

So, as it was shown in \cite{Bjorken&1:AnnPhys:1966}, form factors
of production of resonance with spin $J$ in case of normal parity transition $\gamma^{*}N\to R$
(\ref{N}) have the following threshold behavior:
\begin{equation}
G_{\pm}(Q^2)\sim |\vec{q}|^{J-3/2},
\end{equation}
\begin{equation}
G_{0}(Q^2)\sim \frac{q_0}{|\vec{q}|}\,|\vec{q}|^{J-1/2}.
\end{equation}
In case of abnormal parity transitions (\ref{A}):
\begin{equation}\label{A_G}
G_{\pm}(Q^2)\sim |\vec{q}|^{J-1/2},
\end{equation}
\begin{equation}
G_{0}(Q^2)\sim \frac{q_0}{|\vec{q}|}\,|\vec{q}|^{J+1/2}.
\end{equation}

Special case is transitions $1/2^{+}\to 1/2^{+}$, which are
determined by only two form factors $G_{+}$ and $G_{0}$ ($G_{-}$
corresponds to resonance helicity $3/2$ and thus is absent for
resonances with spin $1/2$). Their threshold behavior for the
transition $1/2^{+}\to 1/2^{+}$ is as follows:
\begin{equation}
G_{+}(Q^2)\sim |\vec{q}|,
\end{equation}
\begin{equation}
G_{0}(Q^2)\sim \frac{q_0}{|\vec{q}|}\,|\vec{q}|^{2}.
\end{equation}
The form factors of transition $1/2^{+}\to 1/2^{-}$ are determined by the expression
(\ref{A_G}) at $J=1/2$, i.e.
\begin{equation}
G_{+}(Q^2)\sim Const,
\end{equation}
\begin{equation}
G_{0}(Q^2)\sim \frac{q_0}{|\vec{q}|}\,|\vec{q}|.
\end{equation}

The behavior of form factors at high $Q^2$ is determined by quark counting rules
\cite{Brodsky&1:PhysRevLett:1973, Matveev&2:LettNuovoCim:1973}, according to which
\begin{equation}
G_{+}(Q^2)\sim Q^{-3},\quad G_{0}(Q^2)\sim Q^{-4},\quad
G_{-}(Q^2)\sim Q^{-5}.
\end{equation}

So, we suggest the expressions for form factors, possessing all the above-mentioned properties,
to be written in the following form:
\begin{gather}\label{FF_N}
\left|G_{\pm}(Q^2)\right|^2=\left|G_{\pm}(0)\right|^2\,
\left(\frac{|\vec{q}|}{|\vec{q}|_{Q=0}}\,\frac{Q_0^{'2}}{Q^2+Q_0^{'2}}\right)^{2J-3}\,
\left(\frac{Q_0^2}{Q^2+Q_0^2}\right)^{m_\pm},\\
\left|G_{0}(Q^2)\right|^2=C^2
\left(\frac{Q^2}{Q^2+Q_0^{''2}}\right)^{2a}
\frac{q_0^2}{|\vec{q}|^2}\,
\left(\frac{|\vec{q}|}{|\vec{q}|_{Q=0}}\,\frac{Q_0^{'2}}{Q^2+Q_0^{'2}}\right)^{2J-1}\,
\left(\frac{Q_0^2}{Q^2+Q_0^2}\right)^{m_{0}}
\end{gather}
for normal parity transitions and
\begin{gather}
\left|G_{\pm}(Q^2)\right|^2=\left|G_{\pm}(0)\right|^2\,
\left(\frac{|\vec{q}|}{|\vec{q}|_{Q=0}}\,\frac{Q_0^{'2}}{Q^2+Q_0^{'2}}\right)^{2J-1}\,
\left(\frac{Q_0^2}{Q^2+Q_0^2}\right)^{m_{\pm}},\\\label{FF_A}
\left|G_{0}(Q^2)\right|^2=C^2
\left(\frac{Q^2}{Q^2+Q_0^{''2}}\right)^{2a}
\frac{q_0^2}{|\vec{q}|^2}\,
\left(\frac{|\vec{q}|}{|\vec{q}|_{Q=0}}\,\frac{Q_0^{'2}}{Q^2+Q_0^{'2}}\right)^{2J+1}\,
\left(\frac{Q_0^2}{Q^2+Q_0^2}\right)^{m_{0}}
\end{gather}
for abnormal parity transitions, where
\begin{equation}
|\vec{q}|=\frac{\sqrt{(M^2-m^2-Q^2)^2+4 M^2 Q^2}}{2M},\quad
|\vec{q}|_{Q=0}=\frac{M^2-m^2}{2M}
\end{equation}
and $m_{+}=3$, $m_{-}=5$, $m_0=4$.

Form factors of transition $1/2^{+}\to 1/2^{+}$ are written as:
\begin{equation}
\left|G_{+}(Q^2)\right|^2=\left|G_{+}(0)\right|^2\,
\left(\frac{|\vec{q}|}{|\vec{q}|_{Q=0}}\,\frac{Q_0^{'2}}{Q^2+Q_0^{'2}}\right)^{2}\,
\left(\frac{Q_0^2}{Q^2+Q_0^2}\right)^{m_{+}},
\end{equation}
\begin{equation}
\left|G_{0}(Q^2)\right|^2=C^2 \left(\frac{Q^2}{Q^2+Q_0^{''2}}\right)^{2a} \frac{q_0^2}{|\vec{q}|^2}\,
\left(\frac{|\vec{q}|}{|\vec{q}|_{Q=0}}\,\frac{Q_0^{'2}}{Q^2+Q_0^{'2}}\right)^{4}\,
\left(\frac{Q_0^2}{Q^2+Q_0^2}\right)^{m_{0}}.
\end{equation}

In the expressions (\ref{FF_N})-(\ref{FF_A}) the quantities $Q_0^2$,
$Q_0^{'2}$, $Q_0^{''2}$ and $a$ are free parameters, which could be determined
by fitting to experimental data. The coefficient $C$
could be determined when the experimental data on the ratio of longitudinal
and transverse cross sections of virtual photon absorption
$R(Q^2)=\sigma_L(Q^2)/\sigma_T(Q^2)$ will become available,
because of the following relation:
\begin{equation}\label{R}
R(Q^2)=\frac{2
\left|G_{0}(Q^2)\right|^2}{\left|G_{+}(Q^2)\right|^2+\left|G_{-}(Q^2)\right|^2}\,.
\end{equation}
By the way, the expression (\ref{R}) together with (\ref{FF_N})-(\ref{FF_A})
allows to determine the behavior of $R$, which may be used for the experimental data
analysis.

The values of form factors at $Q^2=0$ are related to helicity
amplitudes of photoproduction $A_{1/2}$ and $A_{3/2}$, listed in
\cite{PDG:PhysLett:1988},  as follows \cite{Carlson&1:PhysRev:1998}:
\begin{equation}
|G_{+,-}(0)|=e^{-1}\;\sqrt{\frac{M^2-m^2}{m}}\:|A_{1/2,3/2}|\,,
\end{equation}
where $e=\sqrt{4\pi/137}$ is electron charge. Note, that longitudinal form factor
at $Q^2=0$ turns to zero: $G_0(0)=0$.

Substituting expressions (\ref{F1})-(\ref{g2}), written for each particular resonance,
taking into account parity of the transition and using proper expressions for
form factors, to (\ref{StrFun_Sum}), we get the structure
functions in the resonance region:

\begin{multline}\label{F1_res_fin}
F_1(x,Q^2)=\sum_{R}
\frac{m^2}{\pi}\frac{M\Gamma}{(m^2+Q^2(1/x-1)-M^2)^2+M^2\Gamma^2}
\left(\frac{|\vec{q}|}{|\vec{q}|_{Q=0}}\,\frac{Q_0^{'2}}{Q^2+Q_0^{'2}}\right)^{n_{+}}\times\\\times
\left[|G_{+}(0)|^2\left(\frac{Q_0^2}{Q^2+Q_0^2}\right)^{m_{+}}+
|G_{-}(0)|^2\left(\frac{Q_0^2}{Q^2+Q_0^2}\right)^{m_{-}}\right]
\end{multline}

\begin{multline}\label{F2_res_fin}
F_2(x,Q^2)=\sum_R \frac{2m^2
x}{1+4m^2x^2/Q^2}\;\frac{1}{\pi}\;\times
\\ \times
\frac{M\Gamma}{(m^2+Q^2(1/x-1)-M^2)^2+M^2\Gamma^2}\times\\ \times
\left[
\left(\frac{|\vec{q}|}{|\vec{q}|_{Q=0}}\,\frac{Q_0^{'2}}{Q^2+Q_0^{'2}}\right)^{n_{+}}
\left(|G_{+}(0)|^2\left(\frac{Q_0^2}{Q^2+Q_0^2}\right)^{m_{+}}+
|G_{-}(0)|^2\left(\frac{Q_0^2}{Q^2+Q_0^2}\right)^{m_{-}}\right)+\right.\\
+\left. 2C^2 \left(\frac{Q^2}{Q^2+Q_0^{''2}}\right)^{2a}
\frac{q_0^2}{|\vec{q}|^2}\,
\left(\frac{|\vec{q}|}{|\vec{q}|_{Q=0}}\,\frac{Q_0^{'2}}{Q^2+Q_0^{'2}}\right)^{n_{0}}\,
\left(\frac{Q_0^2}{Q^2+Q_0^2}\right)^{m_{0}}\right]
\end{multline}

\begin{multline}
g_1(x,Q^2)=\sum_R \frac{1}{1+4m^2x^2/Q^2}\,\frac{m^2}{\pi}\,
\frac{M\Gamma}{(m^2+Q^2(1/x-1)-M^2)^2+M^2\Gamma^2}\times \\ \times
\left[
\left(\frac{|\vec{q}|}{|\vec{q}|_{Q=0}}\,\frac{Q_0^{'2}}{Q^2+Q_0^{'2}}\right)^{n_{+}}
\left(|G_{+}(0)|^2\left(\frac{Q_0^2}{Q^2+Q_0^2}\right)^{m_{+}}-
|G_{-}(0)|^2\left(\frac{Q_0^2}{Q^2+Q_0^2}\right)^{m_{-}}\right)+\right.\\
+ (-1)^{J-1/2}\,\eta\,\frac{2\sqrt{2}\, m\,x}{Q}\: C\,
\left(\frac{Q^2}{Q^2+Q_0^{''2}}\right)^{a}
\left|G_{+}(0)\right|\times
\\ \times\left.
\frac{|q_0|}{|\vec{q}|}\,
\left(\frac{|\vec{q}|}{|\vec{q}|_{Q=0}}\,\frac{Q_0^{'2}}{Q^2+Q_0^{'2}}\right)^{(n_0+n_{+})/2}\,
\left(\frac{Q_0^2}{Q^2+Q_0^2}\right)^{(m_{0}+m_{+})/2}\right]
\end{multline}

\begin{multline}\label{g2_res_fin}
g_2(x,Q^2)=-\sum_R \frac{1}{1+4m^2x^2/Q^2}\,\frac{m^2}{\pi}\,
\frac{M\Gamma}{(m^2+Q^2(1/x-1)-M^2)^2+M^2\Gamma^2}\times \\ \times
\left[
\left(\frac{|\vec{q}|}{|\vec{q}|_{Q=0}}\,\frac{Q_0^{'2}}{Q^2+Q_0^{'2}}\right)^{n_{+}}
\left(|G_{+}(0)|^2\left(\frac{Q_0^2}{Q^2+Q_0^2}\right)^{m_{+}}-
|G_{-}(0)|^2\left(\frac{Q_0^2}{Q^2+Q_0^2}\right)^{m_{-}}\right)-\right.\\
-
(-1)^{J-1/2}\,\eta\,\frac{Q}{\sqrt{2}\,m\,x}\: C\,
\left(\frac{Q^2}{Q^2+Q_0^{''2}}\right)^{a}
\left|G_{+}(0)\right|\times\\ \times \left.
\frac{|q_0|}{|\vec{q}|}\,
\left(\frac{|\vec{q}|}{|\vec{q}|_{Q=0}}\,\frac{Q_0^{'2}}{Q^2+Q_0^{'2}}\right)^{(n_0+n_{+})/2}\,
\left(\frac{Q_0^2}{Q^2+Q_0^2}\right)^{(m_{0}+m_{+})/2}\right]
\end{multline}
where $n_{+}=2J-3$, $n_{0}=2J-1$ for normal parity transitions and
$n_{+}=2J-1$, $n_{0}=2J+1$ for abnormal parity transitions, and the
sum is over the resonances. We take into account the contributions of the following resonances:
$N(1440)$, $N(1520)$, $N(1535)$, $N(1650)$, $N(1675)$,
$N(1680)$, $N(1700)$, $N(1710)$, $N(1720)$, $N(1990)$,
$\Delta(1232)$, $\Delta(1550)$, $\Delta(1600)$, $\Delta(1620)$,
$\Delta(1700)$, $\Delta(1900)$, $\Delta(1905)$, $\Delta(1910)$,
$\Delta(1920)$, $\Delta(1930)$, $\Delta(1950)$.

The above expressions determine resonance contribution into
nucleon structure functions. It's obvious that the production of resonances
in electron-nucleon scattering is not the only process,
contributing to structure functions. The production of mesons and other hadrons
forms non-resonant background, which also must be taken into account.

Non-resonant background could be parameterized \cite{Niculescu:PhDThesis:1999} as:
\begin{equation}
F_2^{nr}(x,Q^2)=\frac{Q^2}{4\pi^2\alpha}\,\frac{1-x}{1+\frac{4 m^2
x^2}{Q^2}}\:(1+R(x,Q^2))
\sum_{n=1}^{N}{C_n(Q^2)\left[W-W_{th}\right]^{n-1/2}}\,,
\end{equation}
where $N=3$, $C_n$ are fitting coefficients, $W_{th}$ is threshold energy.

\section{Duality}
So finally we obtained nucleon structure functions in the resonance region, dependent on
Bjorken variable and photon virtuality.

Let's consider, following \cite{Bloom&1:PhysRev:1971}, how the
expressions for structure functions (\ref{F1_res_fin})-(\ref{g2_res_fin}),
which contain resonance terms, which have strong dependence on $Q^2$ (because of form factors),
change over in the limit of high $Q^2$ to scaling expressions, which weakly depend on $Q^2$.

We change over to the variable $x'=Q^2/(Q^2+W^2)$, which is related to Bjorken $x$
as follows:
\begin{equation}\label{xprim}
x'=\frac{x\,Q^2}{Q^2+x\,m^2}\,.
\end{equation}

Let's consider the case of infinitely narrow resonances, i.e. $\Gamma\to 0$.
At that, structure function will be nonzero only at those values of $x'$,
at which $W^2=M^2$, i.e.
$x'=Q^2/(M^2+Q^2)$. One may easily see, that the resonance shifts
to the region $x'=1$ as $Q^2$ increases. At that, as experimental data indicate, the resonance follows the scaling curve.
The form of this curve can be evidently derived from the expressions for structure functions
in resonance region by the substitution
$Q^2=M^2\,x'/(1-x')$.

In the case of resonances with finite width the equality $W^2=M^2$
is approximate, which leads to corrections to scaling behavior,
which are proportional to $\Gamma/M$.

In this paper we restrict the consideration to the structure function
$F_2$. It's worth mentioning that structure function $g_1$ is
sensitive, unlike the $F_2$, to longitudinal form factor and
relative values of $G_{+}$ and $G_{-}$, because they are included in
(\ref{g2_res_fin}) as a difference.

In (\ref{F2_res_fin}) we change over to the variable $x'$ and
at high $Q^2$ get:
\begin{multline}\label{scalex}
F_2(x')=\sum_R \frac{2m^2 x'}{\pi M \Gamma}\times\\
\times
\left[
|G_{+}(0)|^2\left(\frac{Q_0^2}{M^2}\right)^{m_{+}}\left(\frac{1-x'}{x'}\right)^{m_{+}}+
|G_{-}(0)|^2\left(\frac{Q_0^2}{M^2}\right)^{m_{-}}\left(\frac{1-x'}{x'}\right)^{m_{-}}+\right.\\
+\left. 2C^2
\left(\frac{Q_0^2}{M^2}\right)^{m_{0}}\left(\frac{1-x'}{x'}\right)^{m_{0}}\right]\,.
\end{multline}
It follows from (\ref{scalex}) that as $x'\to 1$ the structure function
$F_2(x')\sim (1-x')^{m_{+}}$.

Often,~ in the resonance region, Nachtmann variable
$\xi=2x/(1+\sqrt{1+4m^2x^2/Q^2})$ is used (variable $x=Q^2/2m\nu$
is not convenient for the analysis of structure function in resonance region,
because all resonances are densely situated in region $x\sim 1$).
From the inequality $0<x<1$, it's easy to derive that
$0<\xi<Q/m(\sqrt{1+Q^2/4m^2}-Q/2m)$.

The dependence of resonance part of the structure function $F_2$
on Nahtmann variable $\xi$ at various values of $Q^2$ is shown on fig. 2.
Specific values of $Q_0$, $Q_0^{'}$ and $Q_0^{''}$, used to plot fig. 2, are not
significant at this point.
Solid line corresponds to the fit to deep inelastic data for $F_2(\xi)$ \cite{Niculescu&41:PhysRevLett:2000}.

One can see, that $\Delta(1232)$-isobar makes the largest contribution to the structure function.
Non-resonant background in this region
appears to be relatively small, unlike the regions corresponding to
more massive resonances. One can also see that the $\Delta(1232)$ peak
exactly follows the scaling curve as $Q^2$ changes,
which is a manifestation of the duality.

For the comparison with experimental data the parametrization of
non-resonant background contribution must be carried out. However
now this problem has no unique solution.

Taking the non-resonant background into account must
lead to the rise of the structure function
in the region of higher resonances the way it follows the scaling
curve (in varying $Q^2$) in broad region of $\xi$.

\section{Conclusion}
The equality, discovered by Bloom and Gilman, of nucleon structure function $F_2$,
averaged over big enough domain of scaling variable, in resonance region and in deep inelastic
region confirmed the existence of quark-hadron duality,
i.e. the possibility of the description of processes using the language of either only hadronic or
quark degrees of freedom.

Practically it means, that the investigation of nucleon resonances,
for instance in terms of form factors, offers additional
information on nucleon properties in deep inelastic region.

Expressions for form factors of nucleon resonance production, obtained in this work,
taking into account correct threshold behavior and experimental data on resonance decays, may serve
as starting ones for the investigation on nucleon properties in the
language of hadronic degrees of freedom.

Obtained expressions for structure functions allow
to show qualitatively the manifestation of the duality with structure function $F_2$ as an example.

In future, we plan to investigate structure function $g_1$,
which gets essential contribution from the longitudinal form factor.
Numerical verification of our model is also interesting and will be carried out
as new experimental data become available.

\vspace{1cm} Authors thank Laszlo Jenkovszky for helpful discussions.

\newpage
\begin{center} \bf \Large Figures \end{center}
\begin{figure}[h]
\vspace{2cm}
\hspace{6.5cm}
\begin{picture}(20000,20000)
\thicklines
\drawline\photon[\S\REG](5000,15000)[8]
\global\advance\pmidx by 200
\put(\pmidx,\pmidy){$\gamma^{*}$}
\drawline\fermion[\NE\REG](\photonfrontx,\photonfronty)[7000]
\global\advance\fermionbacky by -1200
\global\advance\fermionbackx by -300
\put(\fermionbackx,\fermionbacky){$e^{-}$}
\drawline\fermion[\W\REG](\photonfrontx,\photonfronty)[8000]
\global\advance\fermionbacky by -1000
\global\advance\fermionbackx by 500
\put(\fermionbackx,\fermionbacky){$e^{-}$}
\drawline\fermion[\W\REG](\photonbackx,\photonbacky)[8000]
\global\advance\photonbacky by -50
\drawline\fermion[\W\REG](\photonbackx,\photonbacky)[8000]
\global\advance\photonbacky by 100
\drawline\fermion[\W\REG](\photonbackx,\photonbacky)[8000]
\global\advance\fermionbacky by -1300
\global\advance\fermionbackx by 100
\put(\fermionbackx,\fermionbacky){$N$}
\put(\photonbackx,\photonbacky){\circle*{1500}}
\drawline\fermion[\E\REG](\photonbackx,\photonbacky)[7000]
\global\advance\photonbacky by -500
\drawline\fermion[\E\REG](\photonbackx,\photonbacky)[6500]
\global\advance\photonbacky by 1000
\drawline\fermion[\E\REG](\photonbackx,\photonbacky)[6500]
\global\advance\photonbacky by -500
\global\advance\photonbackx by 7000
\drawline\fermion[\NW\REG](\photonbackx,\photonbacky)[1500]
\drawline\fermion[\SW\REG](\photonbackx,\photonbacky)[1500]
\global\advance\fermionbacky by 700
\global\advance\fermionbackx by 1300
\put(\fermionbackx,\fermionbacky){$R$}
\end{picture}
\vspace{-1.5cm} \caption{Nucleon resonance production in the lowest
order on electromagnetic coupling constant.}\label{Pic10}
\end{figure}
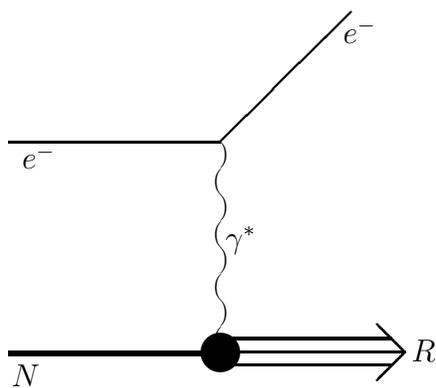

\newpage
\begin{figure}[h]
\vspace{1cm}
\hspace{0.5cm}
\includegraphics[scale=1.5]{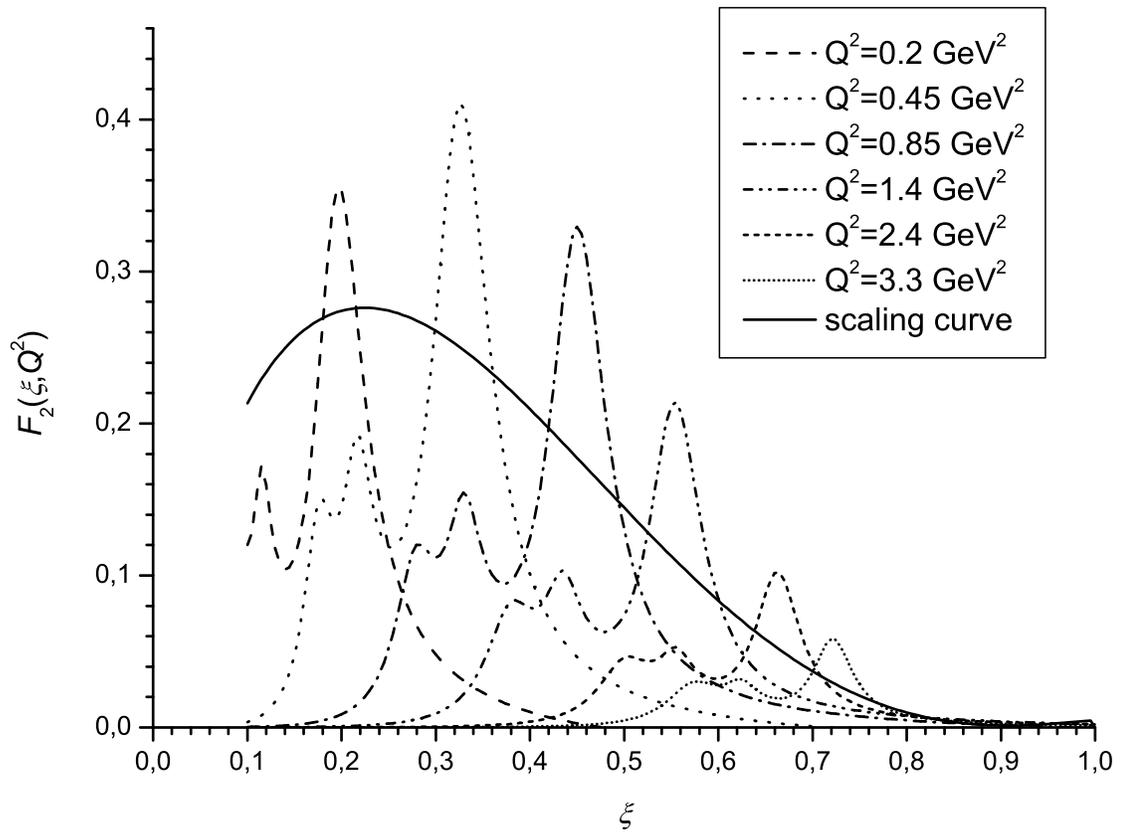}
\vspace{-1cm}
\caption{Resonance contribution to the structure
function $F_2(\xi, Q^2)$ in resonance region.}
\end{figure}

\end{document}